\documentclass[11pt]{article}
\usepackage{latexsym}

\usepackage{amssymb}
\usepackage{amsmath}

 \usepackage{epsfig}
 \usepackage{graphicx}

\newcommand{\be}{\begin{equation}}
\newcommand{\ee}{\end{equation}}
\newcommand{\ba}{\begin{eqnarray}}
\newcommand{\ea}{\end{eqnarray}}
\newcommand{\baa}{\begin{array}}
\newcommand{\eaa}{\end{array}}
\newcommand{\bi}{\begin{itemize}}
\newcommand{\ei}{\end{itemize}}
\newcommand{\edoc}{\end{document}}

\newcommand{\nn}{\nonumber \\}
\newcommand{\nr}[1]{(\ref{#1})}
\newcommand{\la}[1]{\label{#1}}

\newcommand{\rmi}[1]{{\mbox{\scriptsize #1}}}

\newcommand{\fra}[2]{\textstyle{\frac{#1}{#2}\,}}  %%Smaller size!

\newcommand{\bfx}{{\bf x}}

\newcommand{\bfk}{{\bf k}}
\newcommand{\half}{{\textstyle{1\over2}}}

\newcommand{\dd}{\mathrm{d}}

\def\tr{{\rm Tr\,}}

\def\im{{\rm Im\,}}

\def\CL{{\cal L}}

\def\CN{{\cal N}}
\def\CO{{\cal O}}
\def\gsim{\raise0.3ex\hbox{$>$\kern-0.75em\raise-1.1ex\hbox{$\sim$}}}
\def\lsim{\raise0.3ex\hbox{$<$\kern-0.75em\raise-1.1ex\hbox{$\sim$}}}

\begin{document}

\begin{titlepage}
\begin{flushright}
HIP-2010-32/TH\\
%{\Huge DRAFT}
\today\\ %November 2010\\
\end{flushright}
\begin{centering}
\vfill

{\Large{\bf Spatial scalar correlator in strongly coupled hot $\CN=4$ Yang-Mills theory}}

\vspace{0.8cm}

\renewcommand{\thefootnote}{\fnsymbol{footnote}}
% changes the title page footnote counter style to \fnsymbol

K. Kajantie$^{\rm a,b}$\footnote{keijo.kajantie@helsinki.fi},
M. Veps\"al\"ainen$^{\rm a}$\footnote{mikko.vepsalainen@helsinki.fi}
\setcounter{footnote}{0}
% resets the footnote counter after the title page

\vspace{0.8cm}

{\em $^{\rm a}$%
Department of Physics, P.O.Box 64, FI-00014 University of Helsinki,
Finland\\}
{\em $^{\rm b}$%
Helsinki Institute of Physics, P.O.Box 64, FI-00014 University of
Helsinki, Finland\\}
\vspace*{0.8cm}

\end{centering}

\noindent
We use AdS/CFT duality to compute in $\CN=4$ Yang-Mills theory
the finite temperature spatial correlator $G(r)$ of the scalar operator $F^2$,
integrated over imaginary time. The computation is carried out both at
zero frequency and integrating the spectral function over frequencies.
The result is compared with a perturbative computation in finite
$T$ SU($N_c$) Yang-Mills theory.

\vfill \noindent

%\noindent
%PACS numbers:
%11.10.Wx, %        Finite temperature field theory
%11.25.Tq  %	    Gauge/string duality
%05.70.Ce, %        Thermodynamic functions and equations of state

%
%\\
%Keywords:

\vspace*{1cm}

\noindent

%October 2010

\vfill

\end{titlepage}

\section{Introduction}
A project to numerically study spatial correlators $G(r)$ of the scalar and pseudoscalar operators
$\tr F^2$ and $\tr F\tilde F$ in finite temperature gauge field theory
has been initiated in \cite{meyer,iqbalmeyer}. Analytic computations of this correlator
in next-to-leading order QCD perturbation theory have been carried out in \cite{lvv1,lvv2}.
The purpose of this note is to compute this
imaginary-time integrated entirely static
correlator in supersymmetric $\CN=4$ Yang-Mills theory using AdS/CFT duality.
There is a standard framework for this \cite{policastrostarinets,ss,kovtunstarinets,
kovtunstarinets2,romatschkeson},  %%KK
but the computation of full $r$ dependence involves some subtleties in the
subtraction of divergences in the $\omega,k$ plane so that it is perhaps motivated to
report on the details of the computation.

Concretely, we want to compute the finite $T$ correlator, of dimension 7,
\be
G(r)=\int_0^\beta d\tau\,\langle  F^2(\tau,\bfx)F^2(0,{\bf 0})\rangle_T=
\int_0^\beta d\tau \,G(\tau,\bfx), \quad
F^2\equiv \fra14 F_{\mu\nu}^a F_{\mu\nu}^a, \quad r=|\bfx|.
\la{FF}
\ee
This correlator is purely static, $\omega=0$, and we shall determine it by first
evaluating its 3d Fourier transform $G(k)$, of dimension 4, and then transforming to coordinate
space:
\be
G(r)={1\over2\pi^2 r}\int_0^\infty dk\,k \sin(r k)G(k).
\la{gk}
\ee
Hereby it is essential to separate the vacuum part from the finite $T$ part:
\be
G(k)=G_\rmi{vac}(k)+[G(k)-G_\rmi{vac}(k)]=G_\rmi{vac}(k)+G_T(k).
\la{separ}
\ee
We shall include in $G_\rmi{vac}(k)$ all constant terms and evaluate them
analytically. Transforming to
coordinate space they lead to contact terms proportional to $\delta(\bfx)$,
but they have to be subtracted in numerics and their precise value is
important. Furthermore, we show that at large $k$, $G_T(k)$ can be expanded
in powers of $1/k^4$ and evaluate analytically the terms up to order $1/k^8$.

Although the main result can be obtained by taking $\omega=0$, it is also
of interest to see with what happens when the real time
frequency $\omega$ is included. Computing $G(\omega,\bfk)$ and, in
particular, the spectral function $\rho(\omega,\bfk)\equiv \im G_R(\omega,\bfk)$
using the methods of \cite{policastrostarinets,ss,kovtunstarinets,kovtunstarinets2},
the coordinate space Green's function is given by
\be
G(\tau,\bfx)=\int{d^3k\over(2\pi)^3}e^{i\bfx\cdot\bfk}
\int_0^\infty{d\omega\over\pi}\rho(\omega,\bfk)
{\cosh(\fra12\beta-\tau)\omega\over\sinh\fra12\beta\omega}.
\la{spectrep}
\ee
Integrating this over $\tau$ we see that the static correlator is also
obtained by evaluating
\be
G(r)={1\over2\pi^2 r}\int_0^\infty dk\,k \sin(r k)
\int_0^\infty{d\omega\over\pi}{2\rho(\omega,k)\over\omega}.
\la{gkomega}
\ee
We shall show that the same result is obtained via \nr{gk} and \nr{gkomega},
though the path via \nr{gkomega} is much longer.
An essential part of the computation is again correct identification and
elimination of divergences. Furthermore, interesting special nonstatic cases are
the zero momentum temporal correlator $G(\tau,\bfk=0)$ and the equal time
correlator $G(\tau=0,r)$. We shall comment on these at the end. An AdS/CFT
computation of $G(\tau,r)$ has been presented in \cite{iqbalmeyer}.

\section{Equations}
In the standard framework \cite{policastrostarinets,ss,kovtunstarinets,kovtunstarinets2}
and notation, one
takes the background
\ba
ds^2&=&b^2(z)\left(-f(z)dt^2+d\bfx^2+{dz^2\over f(z)}\right),\\
b(z)&=&{\CL\over z},\quad f(z)=1-{z^4\over z_h^4},\quad
{\CL^3\over4\pi G_5}={N_c^2\over2\pi^2},
\la{bg}
\ea
and the scalar field equation therein:
\be
\ddot\phi+\biggl({3\dot b\over b}+{\dot f\over f}\biggr)\dot\phi+
\biggl({\omega^2\over f^2}-{k^2\over f}\biggr)\phi=0,
\la{scaleq}
\ee
where $\phi\equiv\phi(z,\omega,k)\equiv\phi(z,K)$, $k=|\bfk|$ .
Scaling all dimensionful quantities with $z_h$ the equation becomes
\be
\ddot\phi-{3+z^4\over z(1-z^4)}\dot\phi+
\biggl[{\omega^2\over (1-z^4)^2}-{k^2\over1-z^4}\biggr]\phi=0,
\la{eqexpl}
\ee
where
\be
\omega\to z_h\omega={\omega\over\pi T},\quad k\to z_h k={k\over \pi T}.
\la{normal}
\ee
Since
\be
P(z)\equiv {3\dot b\over b}+{\dot f\over f}={d\over dz}\log(b^3f)
\ee
the Wronskian of two any linearly independent solutions of \nr{scaleq} is,
integrating $\dot W/W=-P$,
\be
W(\phi_1,\phi_2)=\phi_1\phi_2'-\phi_2\phi_1'={\bar W_0\over b^3f}=W_0{z^3\over 1-z^4},
\la{Wronsk}
\ee
where $W_0$ is a $z$ independent constant (but will depend on $\omega,k$).

\subsection{The static case, $\omega=0$}\la{subsectstatic}
As explained in the introduction, our goal is to compute the correlator in
the static limit $\omega=0$. In this case it
is a purely Euclidean quantity, and the equation to be solved is simply
\begin{equation}
\ddot\phi-{3+z^4\over z(1-z^4)}\dot\phi-{k^2\over1-z^4}\phi=0\, ,
\label{eqom0}
\end{equation}
with the boundary conditions $\phi(0)=1$ and $|\phi(1)| < \infty$.
While the full solution of this equation is very complicated,
its behavior at large $k$ can be extracted. The leading term will give the
vacuum part of the correlator, which diverges as $k^4$ and has to be subtracted before
Fourier transforming our numerical results into coordinate space. The subleading terms will
enable us to analytically compare the short-distance limit of our result with that
in perturbative QCD \cite{lvv2}, and also to have better control over the numerics.

To find the $k\to\infty$ limit of \nr{eqom0}, the simplest method is to
scale the variable as $y=kz$ and expand in $1/k^4$. The resulting differential
equations are then solved order by order in $k^{-4}$, with the requirement
that the solution stay finite at $z\approx 1$. The leading terms and their expansion at
small $z$ are
\begin{align}
 \phi(z,k) &= \fra12 y^2 K_2(y)
    +{1\over k^4}\cdot{1\over20}\bigl[y^6K_2(y)-y^7K_1(y))\bigr]+\CO(k^{-8}) \nonumber \\
 &\approx 1-\fra14(kz)^2-\fra1{16}k^4z^4(\ln \half k +\ln z+\gamma_E-\fra34)+\fra1{10}z^4 +\CO(z^6).
\la{largekexp}
\end{align}
The constant term $1/10$ will be important in numerics.

This method, while intuitive, is hard to implement beyond this point, as each
new order requires solving an inhomogeneous differential equation
on the whole interval $0\leq y<\infty$ to keep track of the boundary conditions.
In practice it is better to follow the method of
Olver \cite{policastrostarinets,olver}. First we take $u=z^2$ as a new variable and
remove the first derivative by writing $\phi(u)=W(u)\sqrt{u/(1-u^2)}$.
%% MV (see also subsection~\ref{ssec:lc} below).
The resulting equation for $W(u)$ reads
\begin{equation}
  W''(u)=\biggl[ \frac{k^2}{u(1-u^2)} +\frac{3-6u^2-u^4}{4u^2(1-u^2)^2}\biggr]W(u),
\label{eq:static_u}
\end{equation}
where, as always when the variable $u$ is used, $k\equiv k/(2\pi T)$.

Following \cite{olver}, we then rewrite Eq.~(\ref{eq:static_u}) in terms of $\zeta$ and $w(\zeta)$ given by 
\begin{equation}
 \zeta^{1/2} = \int_0^u\!\frac{\dd t}{\sqrt{t(1-t^2)}}, \quad
 \zeta(u) = 4(u+\fra15 u^3 +\fra7{75}u^5)+\CO(u^7),\quad
    W(u) = \left( \frac{\dd\zeta}{\dd u}\right)^{-1/2} w(\zeta)
\ee
and obtain the equation
\be
 w''(\zeta) = \biggl[ {k^2\over4\zeta}+{3\over4\zeta^2} +{\psi(\zeta)\over\zeta} \biggr]w(\zeta),
\end{equation}%%KK
where
\be
\psi(\zeta) = -\fra1{1280}\zeta^3-\fra{17}{665600}\zeta^5+\mathcal{O}(\zeta^7)=
-\fra1{20}u^3-\fra{73}{1300}u^5 +\CO(u^7).
\ee
The solution for $w(\zeta)$ can then be written as a series in inverse powers of $k^2$,
\begin{equation}
 w(\zeta) = \zeta^{1/2}K_2(k\zeta^{1/2}) \sum_{s=0}^\infty \frac{A_s(\zeta)}{k^{2s}}
  -\frac{\zeta}{k} K_3(k\zeta^{1/2}) \sum_{s=0}^\infty \frac{B_s(\zeta)}{k^{2s}}\, ,
\end{equation}
where $A_0=2k^2$ and the other functions $A_s$ and $B_s$ are found using the recursion relations
\begin{align}
 B_s(\zeta) &= -A_s'(\zeta) +\frac{1}{\zeta^{1/2}}\int_0^\zeta\! \frac{\dd v}{v^{1/2}}
    \Bigl[ \psi(v)A_s(v) -\fra{5}{2}A_s'(v)\Bigr], \\
 A_{s+1}(\zeta) &= 2 B_s(\zeta) -\zeta B_s'(\zeta)+\int_0^\zeta \! \dd v\,\psi(v) B_s(v).
\end{align}
The leading term, given by $A_0$ alone, reproduces eq.~(\ref{largekexp}).
For the following terms we note
that for the purposes of computing the correlator, it is sufficient to work
out the power series of $\phi(u)$ around $u=0$ to order $u^2$ (see below).
Using the small-$u$ expansions of $\zeta(u)$ and $\psi(\zeta)$ above we obtain recursively
\begin{align}
A_0 &= \underline{2k^2} &\Rightarrow&& B_0& = -\fra1{35}k^2 u^3 -\fra{667}
{25025} k^2u^5 +\CO(u^7) \nonumber \\
\Rightarrow A_1 &= \fra1{35}k^2 u^3+\fra{1143}{25025}k^2u^5 +\CO(u^7)
    &\Rightarrow&& B_1& = \underline{-\fra{3}{70}k^2u^2} -\fra{5452}
{75075}k^2u^4 +\CO(u^6) \nonumber \\
\Rightarrow A_2 &= \fra{238}{2145}k^2 u^4+\CO(u^6)
    &\Rightarrow&& B_2& = -\fra{136}{715}k^2u^3+\CO(u^5) \nonumber \\
\Rightarrow A_3 &= \fra{136}{715}k^2 u^3+\CO(u^5)
    &\Rightarrow&& B_3& = \underline{ -\fra{204}{715}k^2u^2}+\CO(u^4)\,.
\end{align}
Up to order $\CO(u^2)$, only the underlined terms contribute, finally giving
\begin{align}
 \phi(u) =& 1 -k^2 u -\half k^4 u^2\ln u \nonumber \\
    &+\left[ -k^4\left( \ln k +\gamma_E-\fra34\right) +\frac{1}{10}
    +\frac{3}{35k^4} +\frac{408}{715k^8} +\mathcal{O}(k^{-12}) \right]u^2 +\mathcal{O}(u^3).
    \la{phiuexp}
\end{align}

\subsection{The vacuum contribution}
The $T\to0$ limit amounts to neglecting the $z^4$ terms in
Eq. \nr{eqexpl}. On the $k,\omega$ plane, below the light cone, the solution which
is finite for $z\to\infty$ and normalised to 1 at $z=0$ is
\be
\phi=\fra12 (k^2-\omega^2)z^2 K_2(\sqrt{k^2-\omega^2}z).
\la{K2}
\ee
Continuing across the light cone into $\omega>k$ this becomes
\be
\fra14 i\pi (\omega^2-k^2)z^2H_2^{(1)}(\sqrt{\omega^2-k^2}z).
\la{H2}
\ee
The retarded Green's function following from these \cite{ss} is given
below in \nr{vacG}.

\subsection{The region around the light cone, $\omega\approx k$}\la{ssec:lc}
The numerical results presented later show that the dominant
contribution comes from the region around the light cone. To study this region,
we take in \nr{eqexpl} again $u=z^2$ as a new variable and remove the first derivative
term by writing $\phi(u)=W(u)\sqrt{u/(1-u^2)}$. The equation then becomes
(whenever $u$ is used, $k\equiv k/(2\pi T)$)
\be
W''(u)=\biggl[{-k^2\over(1-u^2)^2}\biggl(u+{\omega^2-k^2\over k^2}{1\over u}\biggr)+
{3\over4u^2}-{u^2\over (1-u^2)^2}\biggr]W(u).
\ee
To find the behavior for large $k$, we scale out $k$ by using $k^{2/3}u$ as
the variable. This leads to the equation
\be
W''(u)=\biggl[-\biggl(1+2\,{\omega^2-k^2\over k^2}\biggr)u+{\omega^2-k^2\over k^{2/3}}
\,{1\over u}+{3\over4u^2}+\CO({1\over k^{8/3}})\biggr]W(u).
\la{nearlc}
\ee
Exactly on the light cone this is simply
\be
W''(u)=\left(-u+{3\over4u^2}\right)W(u),
\ee
which is integrable in terms of Airy functions or Bessel functions of order $2/3$.
The solution corresponding to waves infalling into the black hole
is given by $H_{2/3}^{(1)}$ and the final
solution, returning to $\phi$, is
\ba
\phi(u)&=&k^{2/3}u \,H_{2/3}^{(1)}(\fra23 ku^{3/2})\nn
&=&-i\,3^{2/3}\Gamma(\fra23){1\over\pi}\biggl[1+{(-1)^{1/3}\pi\over 3^{5/6}
\Gamma^2(2/3)}\,k^{4/3}u^2+\CO(u^3)\biggr],
\la{H23}
\ea
where $(-1)^{1/3}=\fra12(1+i\sqrt3)$.
As we shall see concretely, the $u^2$ term here gives directly the Green's function
on the light cone. Near the light cone,
the term $(\omega^2-k^2)/k^2\sim 2(\omega-k)/k$ is very
small and $(\omega^2-k^2)/k^{2/3}\sim2k^{1/3}(\omega-k)$. Thus one predicts that the
outcome will depend mainly on the combination $k^{1/3}(\omega-k)$. We shall confirm
this numerically (see Fig.~\ref{overpeak}).

\subsection{Method of numerical solution}
The computation of the correlator
begins by numerically finding solutions $\phi_h(z,K),\,\,K=(\omega,k)$
of \nr{eqexpl} representing infalling waves,
%$\exp[-i\omega(t-\fra12(\arctan(z)+{\rm atanh}(z)))]
$\sim\exp(-i\omega t)(1-z)^{-i\omega/4}$.
These satisfy $\phi_h^*(z,K)=\phi_h(z,-K)$. For $\omega=0$ the situation is particularly
simple, one can choose a solution normalised to 1, the other is logarithmically
divergent.

Because of the $1/(1-z^4)$ factor, the integration cannot be started exactly at
$z=1$. One then expands the solution around $z=1$. The expansion starts
\ba
\phi_h(z)&=&(1-z)^{-\frac{i \omega}{4}} \left[1+\frac{(1-z) \left(4 k^2-3 \omega
   (\omega-2 i)\right)}{16-8 i \omega}
\right.\nn&&\left.-\frac{(1-z)^2 \left(16
   k^4-24 k^2 \omega^2+\omega \left(9 \omega^3+2 i
   \omega^2+48 \omega+32 i\right)\right)}{128 \left(\omega^2+6
   i \omega-8\right)}+...\right];
\la{infalling}
\ea
up to 20 terms were used. One then expands this solution in terms of the two
independent solutions at the boundary $z=0$, the unnormalisable solution
$\phi_u(z,K)$ and the normalisable solution $\phi_n(z,K)$:
\be
\phi_h(z,K)=A(K)\phi_u(z,K)+B(K)\phi_n(z,K).
\la{hAB}
\ee
The expansions start (up to order $z^{40}$ were used) as
\ba
\phi_u(z,K)&=&1+\frac{1}{4} z^2 \left(\omega^2-k^2\right)+\frac{1}{288} z^6
   \left(k^6-3 k^4 \omega^2+3 k^2
   \left(\omega^4-8\right)-\omega^6\right)+\CO(z^8)\nn&&
   -{(\omega^2-k^2)^2\over16}\log(z)\phi_n(z,K),
\la{phiu}
\ea
\ba
\phi_n(z,K)&=&z^4\left[
1+\frac{1}{12} z^2 \left(k^2-\omega^2\right)+\frac{1}{384}
   z^4 \left(k^4-2 k^2 \omega^2+\omega^4+192\right)\right.\la{phin}
   \\&&\left.+\frac{z^6\left(k^6-3 k^4 \omega^2+3 k^2 \omega^4+1344
   k^2-\omega^6-1728 \omega^2\right)}{23040}\right]+\CO(z^{12}).
\ea
Their Wronskian is $\phi_u\dot\phi_n-\phi_n\dot\phi_u=4z^3/(1-z^4)$. Similarly,
from \nr{hAB}, $W(\phi_h,\phi_n)=4A(K)z^3/(1-z^4)$ and $W(\phi_h,\phi_u)=-4B(K)z^3/(1-z^4)$.

For the Green's function one needs \cite{ss}, expanding near $z=0$,
\be
{1\over z^3}{\dot\phi_h(z,K)\over \phi_h(z,K)}=
{\omega^2-k^2\over2z^2}-{(\omega^2-k^2)^2\over16}(4\log z+3)+4\cdot {B(K)\over A(K)}+\CO(z^2).
\la{keyeqexp}
\ee
The first two real terms are neglected as contact terms (and they anyway vanish on
the light cone) and
the result for the Green's function is, including $1/(16\pi G_5)$ from the gravity action
\be
G(K)={\CL^3\over4\pi G_5}{B(K)\over A(K)},
\la{keyeq}
\ee
where
\be
{B(K)\over A(K)}=-
{\phi_h\dot\phi_u-\phi_u\dot\phi_h\over \phi_h\dot\phi_n-\phi_n\dot\phi_h}.
\la{Gfin}
\ee
Since both Wronskians are $\sim z^3/(1-z^4)$, the ratio is independent of
$z$ and could, in principle, be evaluated at any $z$.
In practice, the $z$-independence is
limited by how many terms are included in the small-$z$ expansions of
$\phi_u,\,\phi_n$.
Note that the dimensions of $B,\phi_n$ are $+4,-4$; $A,\phi_u$ are dimensionless.
The neglect of the divergent real terms in the expansion \nr{keyeqexp} may seem
somewhat surprising,
without counter terms the result \nr{keyeq} only obviously holds for the imaginary
part. We find that it also reproduces the real part correctly when it can
be analytically computed.

As a first application \cite{ss}, from the properly analytically continued
vacuum solutions \nr{K2} and \nr{H2} one finds that
\be
G_\rmi{vac}(\omega,k)=-{(\omega^2-k^2)^2\over32}\biggl[\log{-(\omega^2-k^2)\over4}+
2\gamma_E-\fra32\biggr].
\la{vacG}
\ee
Here and until further notice we omit the normalisation factor
$\CL^3/(4\pi G_5)=N_c^2/(2\pi^2)$.

\begin{figure}[!tb]
\begin{center}

\vspace{0.5cm}
\includegraphics[width=0.49\textwidth]{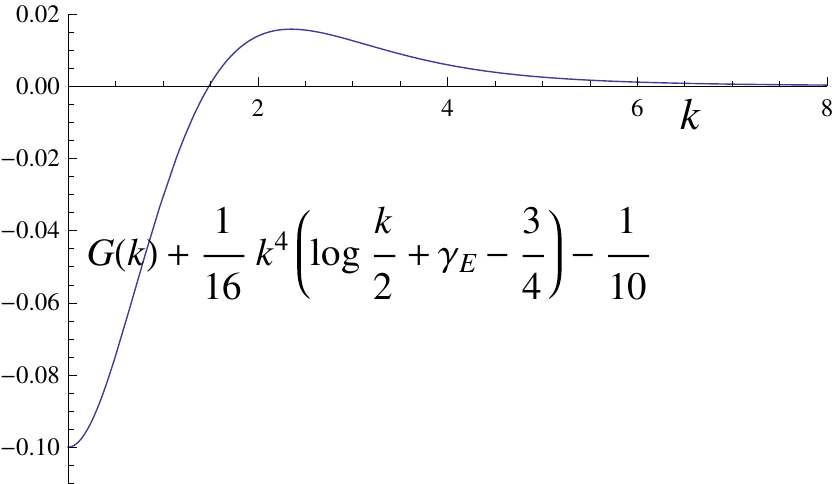}\hfill
\includegraphics[width=0.49\textwidth]{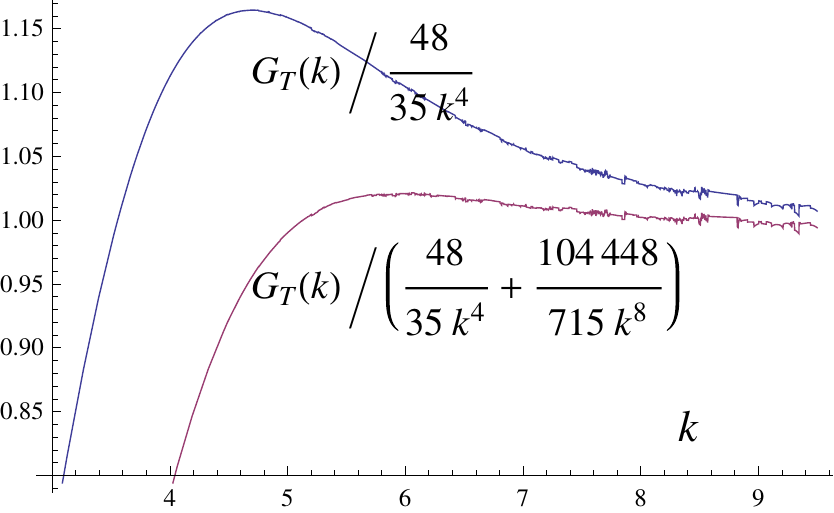}
\end{center}

\caption{\small Left panel: Numerically computed finite $T$ part $G_T(k)$ of the correlator,
the vacuum part \nr{GTk} is
subtracted. Right panel: Check of the analytic expressions \nr{k4k8} for the large-$k$
behavior of $G_T(k)$. The constants here were originally determined numerically, only
later confirmed analytically. }
\la{Gkstatic}
\end{figure}

\section{Numerical results for $G(k)$}
\subsection{The static case $\omega=0$}
Comparing Eqs. \nr{phiuexp} and \nr{hAB} and changing from $k\equiv k/(2\pi T)$
back to $k\equiv k/(\pi T)$ one sees that $A(k)=1$ and
\be
B(k)=-\fra1{16}k^4\left[\log\fra12k+\gamma_E-\fra34\right]+\fra1{10}+\CO(k^{-4}).
\ee
There thus is a diverging vacuum contribution plus a constant term $1/10$. Including
this in $G_\rmi{vac}$ we write
\ba
G(k)&=&G_\rmi{vac}(k)+[G(k)-G_\rmi{vac}(k)]=G_\rmi{vac}(k)+G_T(k),\nn
G_\rmi{vac}(k)&=&-\fra1{16}k^4\left[\log\fra12k+\gamma_E-\fra34\right]+\fra1{10}.
\la{GTk}
\ea
Furthermore, the large-$k$ behavior of $G_T$ is
\be
G_T(k)={48\over35 k^4}+{104448\over 715k^8}+\CO(k^{-12}).
\la{k4k8}
\ee

In the numerical integration of \nr{eqom0} the initial condition is simply $\phi_h(1,0,k)=1$,
the other solution diverges logarithmically. Because of the $1/(1-z^4)$ factor the integration
is started from $z=1-\epsilon'$
($\epsilon'=0.2$), correcting the initial conditions of $\phi_h$ and $\dot\phi_h$ by the
expansion \nr{infalling} (up to $(1-z)^{20}$ was used). Eq. \nr{Gfin} is then evaluated at some
$z=\epsilon$ ($\epsilon=0.2$) using the expansions \nr{phiu} and \nr{phin} (up to
$z^{40}$ was used).
The finite $T$ part obtained after subtracting the vacuum part in \nr{GTk}
is plotted in Fig. \ref{Gkstatic}. The figure also shows how well its
large $k$ behavior in Eq. \nr{largekexp} is reproduced. One sees that
this expansion converges rapidly, already
at $k=6$ the error is $<1\%$. What is important is that there are no terms
decreasing more slowly.

Fig. \ref{Gkstatic} gives the quantities in dimensionless units. If $G_T$ and $k$ are
in physical units, Fig.~\ref{Gkstatic} plots $G_T(k)/(\pi T)^4$ vs $k/(\pi T)$.

\subsection{The case with $\omega\not=0$}
The static correlator $G(r)$ is obtained by Fourier transforming $G(k)$ in the
previous subsection, but it may be illuminating to see how the same result is
obtained with full $\omega$ dependence, using Eq. \nr{gkomega}.

To begin with, from \nr{vacG} one finds that ($\omega>0$)
\be
\im G_\rmi{vac}\equiv\rho_\rmi{vac}(\omega)={\pi\over32}(\omega^2-k^2)^2\Theta(\omega-k).
\la{vacrho}
\ee
Fig. \ref{vacfig} shows how well the numerics approaches this vacuum spectral
function. This divergence will be subtracted in what follows.

\begin{figure}[!t]
\begin{center}

%\vspace{-0.8cm}
\includegraphics[width=0.5\textwidth]{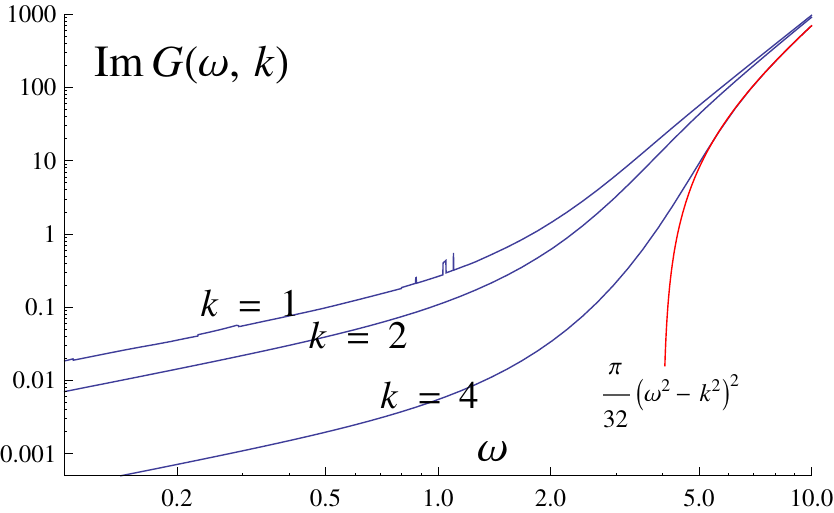}
\end{center}

\caption{\small Data for $\rho(\omega)\equiv{\rm Im} G(\omega,k)$ for
$k =1,2,4$ plotted vs $\omega$,
The approach to the vacuum spectral function \nr{vacrho} is shown by the red curve for $k=4$. }
\la{vacfig}
\end{figure}

\begin{figure}[!tb]
\begin{center}

\vspace{0.5cm}
\includegraphics[width=0.45\textwidth]{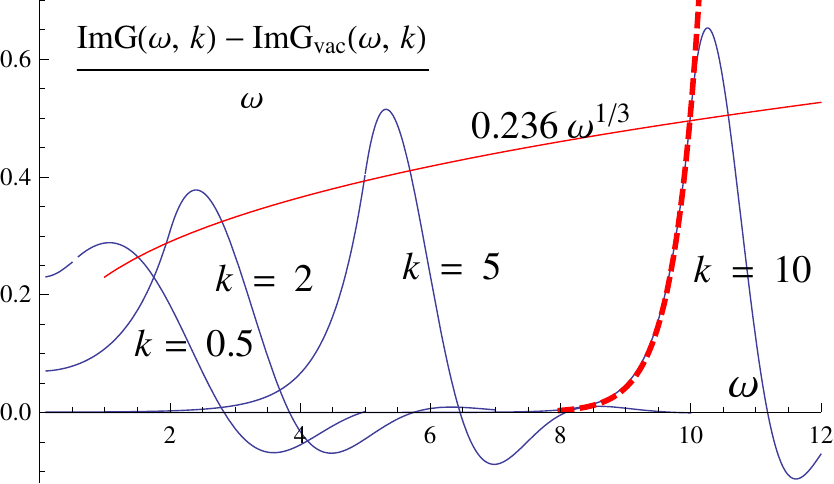}\hfill
\includegraphics[width=0.45\textwidth]{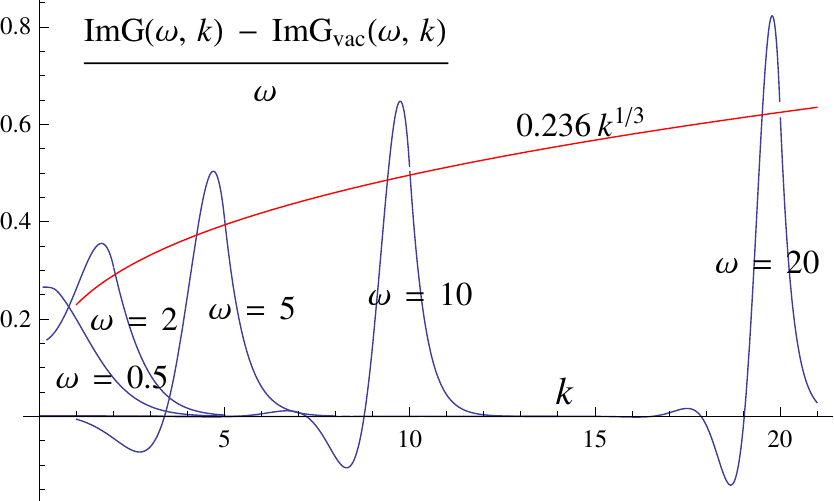}
\end{center}

\caption{\small Data for $[\rho(\omega,k)-\rho_\rmi{vac}(\omega,k)]/\omega$,
$\rho_\rmi{vac}=\pi/32\cdot(\omega^2-k^2)^2\Theta(\omega-k)$, the
left panel plots fixed values of $k$ as a function of $\omega$, the right
panel the opposite. The red curve shows the result \nr{LCresult} exactly on the
light cone, $\rho(k,k)/k$. The red dotted curve is an exponential
fit $0.51 e^{2.5(k-10)}$. }
\la{fig:Gomegak}
\end{figure}

After vacuum subtraction one finds that $\rho-\rho_\rmi{vac}$ is non-zero essentially
close to the light cone. The solution for $\phi$ exactly on the light cone
is given by the Hankel function in \nr{H23}. The Green's function is evaluated from
\nr{Gfin} or directly by computing $2/u \cdot\phi'(u)/\phi(0)$ with the result
\be
G(k,k)= (1+i\sqrt3){\Gamma(1/3)\over8\cdot6^{1/3}\Gamma(2/3)}\,k^{4/3}
+\CO(1)
\approx(0.136092+i\,0.235718)\,k^{4/3}.
\la{LCresult}
\ee
In transforming \nr{H23} one must remember that there $k\equiv k/(2\pi T)$ while in the
numerics $k\equiv k/(\pi T)$. Also note that $\Gamma(\fra23)\Gamma(\fra13)=2\pi/\sqrt3$.
Fig. \ref{fig:Gomegak} plots the vacuum-subtracted imaginary part either
as a function of $\omega$ for various values of $k$ or
as a function of $k$ for various values of $\omega$ and one sees that the analytic form
\nr{LCresult} works very well at $\omega=k$ (note that this is not the top of the
curves), even down to $k\gsim2$.
The left panel of Fig. \ref{fig:Gomegak}
shows how the decrease off the peak is roughly exponential \cite{ss}.
For the real part (not shown) the agreement is similar, though the $k^{4/3}$ behavior
sets in at somewhat larger values of $k\gsim10$.

\begin{figure}[!h]
\begin{center}

\includegraphics[width=0.7\textwidth]{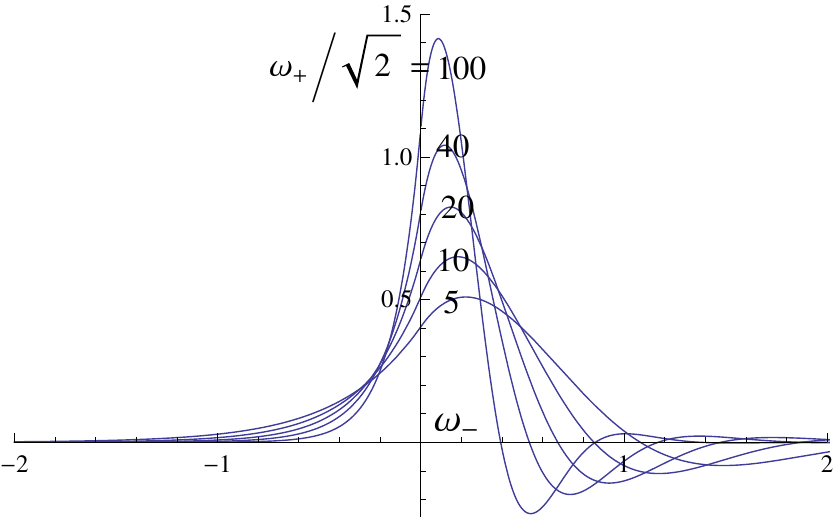}
\end{center}

\vspace{-0.6cm}
\caption{\small Data for $[\rho(\omega,k)-\rho_\rmi{vac}(\omega,k)]/\omega$ when
crossing the light cone in a perpendicular direction,
i.e., in the direction of $\omega_-$, $\omega_\pm=(\omega\pm k)/\sqrt2$, plotted
for increasing values of $\omega_+$. Note how the peak gets narrower with
increasing $\omega_+$, the curves are of the form $k^{1/3}\,F(k^{1/3}\omega_-$).
The real part has a similar structure, but the positive peak is at $\omega_-<0$
and the dips at $\omega_->0$ are deeper.}
\la{overpeak}
\end{figure}
\begin{figure}[!h]
\begin{center}

\vspace{0.5cm}
\includegraphics[width=0.45\textwidth]{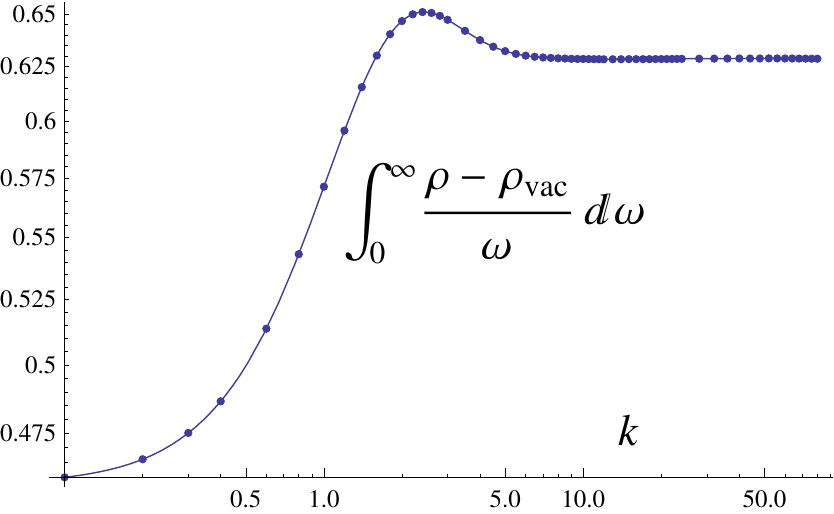}\hfill
\includegraphics[width=0.45\textwidth]{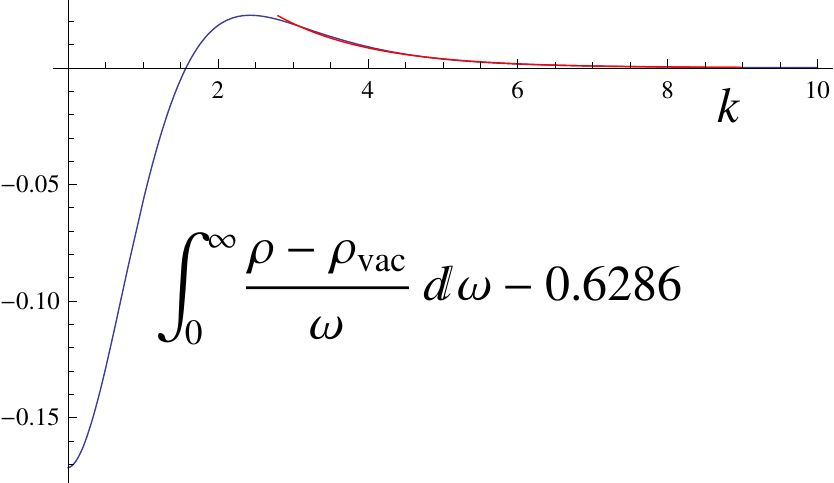}
\end{center}

\caption{\small Data for the integral \nr{omegaint} as a function of $k$.
The limits for $k\to0$ and $k\to\infty$ are $3\pi/20$ and $4\pi/20$.%%KK
The right panel has the numerically obtained asymptotic value $0.6286$ subtracted.}
\la{intk}
\end{figure}

To study in more detail the region around the light cone,
Fig. \ref{overpeak} shows the variation when
one crosses the light cone perpendicularly to it, in the direction of
$\omega_-=(\omega-k)/\sqrt2$ at fixed $\omega_+=(\omega+k)/\sqrt2$.
One sees that the prediction made
after Eq. \nr{H23} is confirmed: the quantity in Fig.~\ref{overpeak} depends essentially on the
combination $k^{1/3}(\omega-k)$, it is of the form $k^{1/3}F(k^{1/3}\omega_-)$,
where $F(0)$ can be read from \nr{LCresult}.
Thus while the peak
height increases, its width decreases. A consequence of this is that the
integral over the peak,
\be
\int_0^\infty d\omega\,{\rho(\omega,k)-\rho_\rmi{vac}(\omega,k)\over\omega},
\la{omegaint}
\ee
approaches a constant at large $k$, see Fig. \ref{intk}. The accuracy is such that
within the range 10...80 the last digit in 0.6286 varies between 3...8.
Subtracting
this constant one obtains the right panel of Fig. \ref{intk}. After multiplication
by $2/\pi$ this curve is, within numerical accuracy, the same as in the left panel of
Fig. \ref{Gkstatic}. One has, along a rather roundabout way on the $k,\omega$ plane,
shown that Eqs. \nr{gk} and \nr{gkomega} lead to the same result.

That this numerically obtained constant is actually $\pi/5$ can be derived as follows
by first deriving its value at $k=0$.
A properly subtracted dispersion relation is (at any $k$)
\be
G(\omega)-G_\rmi{vac}(\omega)-\bigl[G(\omega\to\infty)-G_\rmi{vac}(\omega\to\infty)\bigr]
=\int_{-\infty}^\infty {d\,\omega'\over\pi}
{\rho(\omega')-\rho_\rmi{vac}(\omega')\over \omega'-\omega}.
\la{disprel}
\ee
For $k=0$ one can work out the $\omega\to\infty$ limit of $G(\omega,k=0)$ exactly
as the $k\to\infty$ limit of $G(\omega=0,k)$ was worked out in subsection \ref{subsectstatic},
with the result
$G(\omega\to\infty)-G_\rmi{vac}(\omega\to\infty)=-3/10+\CO(1/\omega^4)$ \cite{romatschkeson}.
This constant $-3/10$ is the analogue of the constant 1/10 in Eq.~\nr{largekexp}. If one
now takes $\omega=0$ in \nr{disprel}, the LHS leaves just the constant $3/10$ and the
RHS is $2/\pi\times$ \nr{omegaint} at $k=0$: the value of \nr{omegaint} at $k=0$ is
$3\pi/20$. Since the $k$ dependence has to be the same as that in the static case,
Fig.~\ref{Gkstatic}, the value of \nr{omegaint} at $k=\infty$ is $4\pi/20$.

\section{$G(r)$}
\subsection{The static case $\omega=0$}
To finally get the correlator in coordinate space we have to compute the integral
(the normalisation $\CL^3/(4\pi G_5)$ will be appended later)
\be
G(r)=\int{d^3k\over(2\pi)^3} e^{i\bfk\cdot\bfx}\left[G_\rmi{vac}(k)+G_T(k)\right].
\ee
In the Fourier transform of $G_\rmi{vac}$ all terms but the $\log k$ term in
\nr{GTk} are contact terms, proportional to $\delta(\bfx)$ or derivatives thereof,
and can be neglected. Using
\be
\log k=\int_0^\infty {dt\over t}\left(e^{-t}-e^{-kt}\right)
\ee
the $\log k$ term becomes
\be
\int{d^3k\over(2\pi)^3} e^{i\bfk\cdot\bfx}G_\rmi{vac}(k)=-\infty\,{\rm contact\,\,term}+
{1\over32\pi^2r}\int_0^\infty{dt\over t}\int_0^\infty dk\,k^5\sin(r k)e^{-kt}=
{15\over 8\pi r^7}.
\ee
If we introduce $\CL^3/(4\pi G_5) = N_c^2/(2\pi^2)$, as is
appropriate for $\CN=4$ conformal field theory, and restore physical units by
appropriate powers of $z_h$, the final result can be written as
\be
{G(r)\over N_c^2(\pi T)^7}={15\over16\pi^3\bar r^7}+
{1\over4\pi^4\bar r}\int_0^\infty dk\,k \sin(\bar rk)G_T(k),\quad \bar r\equiv
\pi Tr.
\la{res}
\ee

Comparing with the next-to-leading order computation in SU($N_c$) Yang-Mills theory
\cite{lvv2} one notes that the leading UV terms\footnote{The leading diagram of the
$F^2$ correlator is reduced to scalar master integrals in Eq. (3.1) of \cite{lvv1}.
The master integrals are evaluated in Eq. (A.16) and the final Fourier transformation
from $G(p)$ to $G(r)$
is carried out using Eq. (5.1)} are the same if one corrects
for an obvious factor of 16. This arises since \cite{lvv1} studies
correlators of $F_{\mu\nu}^aF_{\mu\nu}^a$ while in AdS/CFT duality the scalar couples to
the operator $\fra14 F^2$ with the factor $\fra14$.

The finite $T$ part of the full result in
\nr{res}, multiplied by 16, is compared with the NLO computation in
\cite{lvv2} in Fig. \ref{comp}; the leading $1/r^7$ terms are the same.
One observes that in the
finite $T$ part there is a similar region of negative
correlator around $\pi Tr=1$, but there is one crucial difference, the
QCD result contains terms $\sim (\epsilon+p)/r^3$ and $(\epsilon-3p)/r^3$
while there are no terms $\sim 1/r^3$ nor other terms diverging for $r\to0$
in the finite $T$ AdS result.
The NLO QCD result turns positive at $\pi T r\approx2.6$, we do not see this in the
AdS result.

The relative magnitudes of the vacuum and finite $T$ parts of \nr{res} are
compared in Fig. \ref{comp2}. One sees that the vacuum part dominates for
$\pi Tr<1$, the finite $T$ part grows in relative importance for $1<\pi Tr<4$ and for
$\pi Tr>4$ both terms essentially cancel each other.

\begin{figure}[!t]
\begin{center}

\vspace{0.5cm}
\includegraphics[width=0.49\textwidth]{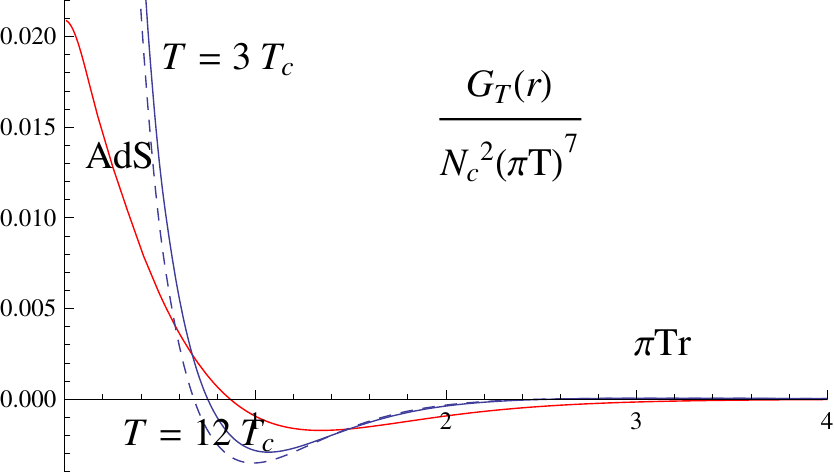}\hfill
\includegraphics[width=0.49\textwidth]{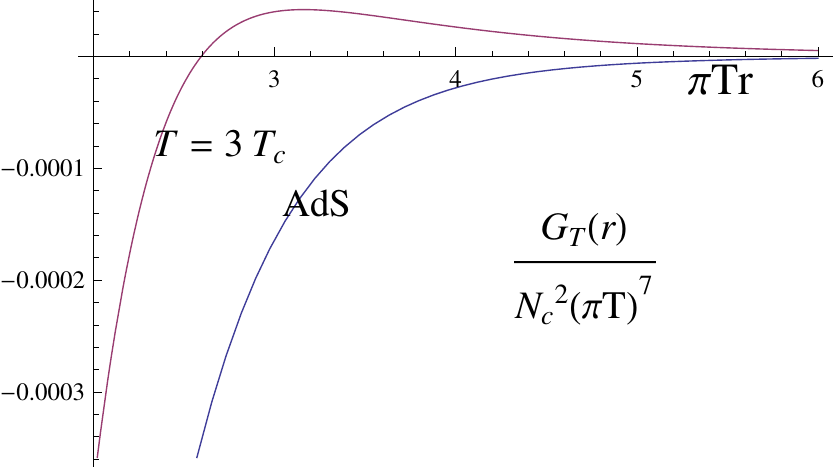}
\end{center}

\caption{\small Comparison of the result $G_T$ for the finite $T$ part
in \nr{res} (multiplied by 16 to get a correlator
of $F^2$) with the
NLO computation in SU($N_c$) Yang-Mills theory in \cite{lvv2} for $T=3T_c$ and
$T=12T_c$ (dashed). The right panel shows the large $k$ region. The leading
$1/r^7$ divergence at small $r$
(not shown) is the same for the two computations. }
\la{comp}
\end{figure}

\begin{figure}[!t]
\begin{center}

\vspace{0.5cm}
\includegraphics[width=0.4\textwidth]{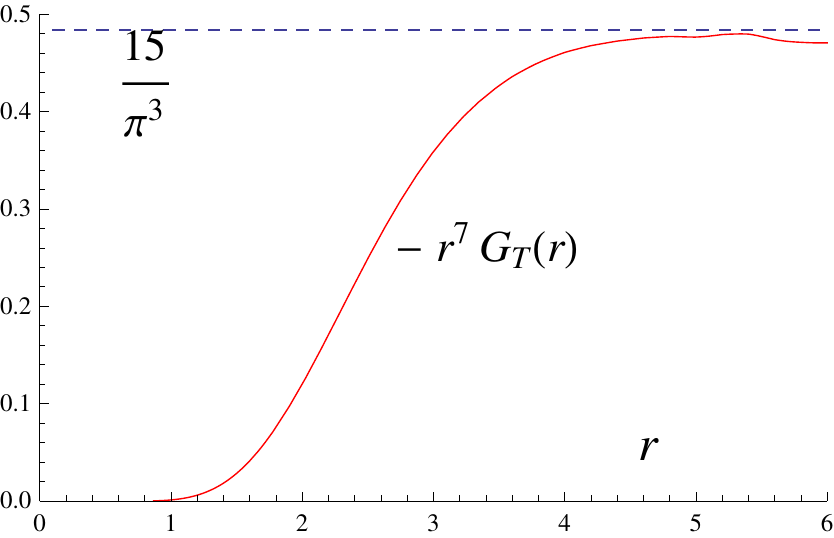}\hfill

\end{center}

\caption{\small Comparison of the relative magnitudes of the vacuum and
finite $T$ parts of \nr{res} (multiplied by 16 to get a correlator
of $F^2$). Both terms are multiplied by $r^7$, $r\equiv\pi Tr$.}
\la{comp2}
\end{figure}

\subsection{The case with $\omega\not=0$}
In this case one can write, from Eq. \nr{gkomega},
\ba
G(r)
&=&{1\over2\pi^2 r}\int_0^\infty dk\,k \sin(r k)
\biggl[\int_{-\infty}^\infty{d\omega\over\pi}{\rho_\rmi{vac}(\omega,k)\over\omega}+
\int_{-\infty}^\infty{d\omega\over\pi}{\rho(\omega,k)-\rho_\rmi{vac}(\omega,k)\over\omega}\biggr] \nn
&=&{1\over2\pi^2 r}{2\over\pi}
\int_0^\infty dk\,k \sin(r k) \times \nn 
&&\hspace{6em}\biggl[\int_k^\infty{d\omega\over\omega}{\pi\over32}
(\omega^2-k^2)^2e^{-\tau\omega}+\int_0^\infty d\omega{\rho-\rho_\rmi{vac}\over\omega}-0.6286+0.6286
\biggr]\nn
&=&{1\over\pi^3 r}\biggl\{{15\pi^2\over8r^6}+
\int_0^\infty dk\,k \sin(rk)\biggl[\int_0^\infty d\omega{\rho-\rho_\rmi{vac}\over\omega}-0.6286\biggr]
\biggr\}+{\rm const}\cdot\delta(\bfx).
\la{finalint}
\ea
Here the Fourier transform of the vacuum contribution is computed by introducing a convergence
improving factor $e^{-\tau \omega}$. The vacuum integrals over $\omega$ and $k$ can
then be done analytically and the limit $\tau\to0$ leads to the expression
in \nr{finalint} above. Since the square bracket in the integrand in the RHS,
after multiplication with $2/\pi$,
coincides with $G_T(k)$, this result is the same as in \nr{res}.

\section{$G(\tau,\bfk=0)$ and $G(\tau=0,r)$}
Various $\tau$ dependent correlators can also be computed using the spectral
representation \nr{spectrep}. A particularly simple case is
the $\tau$ dependent correlator
at zero spatial momentum, obtained by summing over spatial volume.
Using the spectral representation
\nr{spectrep} and the explicit form \nr{vacrho} one has, within
the range $0<\tau<1/(2T)$,
\ba
G(\tau,\bfk=0)&=&\int_0^\infty d\omega\biggl[\fra1{32}\omega^4+
\fra1\pi(\rho(\omega,0)-\rho_\rmi{vac}(\omega,0))\biggr]
{\cosh[(1-2T\tau)\fra\pi2\omega]\over \sinh\fra\pi2\omega}\nn
&=&{3\over4\pi^5}[\zeta(5,T\tau)+\zeta(5,1-T\tau)]+G_2(T\tau),
\la{Gtau0}
\ea
where one has noted that in the dimensionless units used here $\beta\omega\to\pi\omega$
and where $G_2$ has to be integrated numerically. The integrand is similar to the $k=0.5$
curve in Fig. \ref{fig:Gomegak}, multiplied by $\omega$. For small $x$,
\be
\zeta(5,x)+\zeta(5,1-x)={1\over x^5}+2\zeta(5)+30\zeta(7)x^2+\CO(x^4).
\ee
Physical dimensions of $G$ are restored by multiplying by $(\pi T)^5$ so that the
first term produces the $T$ independent UV divergence $3/(4\tau^5)$. Again,
there are no further divergent terms. One finds that
the component $G_2$ is numerically insignificant even at $T\tau=\fra12$, where it is
largest relative to the $\zeta$ function terms.

A more complicated case is $G(\tau,r)$. We give, for completeness, only the part
obtained by inserting to \nr{spectrep} the vacuum spectral function \nr{vacrho}:
\be
G_\rmi{vac}(\tau,r)={6\over\pi^2r^8}\sum_0^\infty\biggl\{
{1\over[\pi^2(n+T\tau)^2/r^2+1]^4}+{1\over[\pi^2(n+1-T\tau)^2/r^2+1]^4}\biggr\}.
\la{Gtaur}
\ee
The sum can be expressed in terms of derivatives of the gamma function, but this is not
very illuminating. For equal $\tau$ correlators one has
\ba
&&G_\rmi{vac}(\tau=0,r)={6\over\pi^2r^8}
\left[
\frac{1}{24} r^4 \text{csch}^4(r)+\frac{1}{12} r^4 \coth ^2(r) \text{csch}^2(r)+\frac{1}{4} r^3
   \coth (r) \text{csch}^2(r)+\right.
   \nn&&\left.\frac{5}{16} r^2 \text{csch}^2(r)+\frac{5}{16} r \coth (r)
\right],
   \ea
where the quantity in the brackets $=1+r^8/4725+\CO(r^{10})$. Including the factor $N_c^2/(2\pi^2)$
and multiplying by 16 to get correlators of $F^2$, the leading singular term $\sim48N_c^2/(\pi^4r^8)$
is seen to agree with that in \cite{lvv1}.

\section{Conclusions}
Motivated by needs of lattice work \cite{meyer},
we have in this article computed the $r$ dependence of the $\tau$ integrated finite
temperature correlator $\langle F^2(\tau,\bfx)F^2(0,{\bf 0})\rangle$ and compared it
with the same in SU($N_c$) Yang-Mills theory \cite{lvv2}. Our computation based on
AdS/CFT duality applies to $\CN=4$ conformal strongly coupled theory and there is
no reason for the results to coincide. The leading UV vacuum part $\sim1/r^7$, $r\to0$,
coincides since it is independent of the coupling.
However, the leading finite $T$ parts differ, in the QCD-like case there are terms
$\sim(\epsilon+p)/r^3$ and $\sim(\epsilon-3p)/r^3$, while for AdS/CFT we find no terms
$\sim T^4/r^3$ nor other finite $T$ terms diverging for $r\to0$.
$G_T$ is just some function $G_T(\pi Tr)$, finite for $r\to0$. In the range
$r\sim{\rm few}\,\,1/(\pi T)$ both finite $T$ correlators are negative. For
$r>4/(\pi T))$ the AdS finite $T$ correlator essentially cancels the vacuum
one.

The computation was carried out both taking $\omega=0$ from the outset, but also
by integrating the weighted spectral function over $\omega$. These results coincide
due to observed structure near the light cone, $\omega=k$. It would be interesting to
have analytic control of observed scaling as a function of $k^{1/3}(\omega-k)$.

An obvious task for the future would be carrying out a similar computation for
holographic QCD models \cite{kiri1,kiri2} breaking conformal invariance.
This will also shed light on the difference between
$\langle F^2(x)F^2(0)\rangle$ and $\langle F^2(x)\tilde F^2(0)\rangle$
correlators.

\vspace{1cm}
{\it Acknowledgements}.
KK thanks J. Alanen, Sean Nowling, Aleksi Vuorinen and, in particular,
Jorge Casalderrey-Solana for discussions and advice.
The work of MV has been supported by Academy of Finland, contract no. 128792.

\end{document}